# Development of the Miniaturised Endoscope Holder LER (Light Endoscope Robot) for Laparoscopic Surgery

LONG Jean-Alexandre (1), CINQUIN Philippe (2), TROCCAZ Jocelyne (2), VOROS Sandrine (2), DESCOTES Jean-Luc (1), BERKELMAN Peter (2), LETOUBLON Christian (3),  RAMBEAUD Jean-Jacques (1)

(1) Urological Surgery and Renal Transplantation Unit, Grenoble University Hospital Centre – 30043 Grenoble France
(2) TIMC-GMCAO Laboratories (UMR CNRS 5525) Grenoble
(3) General and Digestive Surgery Unit, Grenoble University Hospital Centre, 30043 Grenoble cedex 9- France

Mailing Address:

Dr Jean-Alexandre LONG

Urological Surgery and Renal Transplantation Unit

Michallon University Hospital Centre

38043 Grenoble cedex 9

e-mail: JALong@chu-grenoble.fr

Fax: (33-4) 76 76 56 11

**Abstract**

**Introduction**: Our presentation focuses on our 1st experiments with an innovatively-designed robot endoscope holder for laparoscopic surgery, which has a limited size and cost.

**Materials and Methods**: A compact robot (LER) placed on the patient's skin that can be used in the lateral and dorsal supine position was tested on cadavers and laboratory pigs in order to allow successive modifications. The current control system is based on voice recognition. The amplitude of vision is 360 degrees with an angle of 160 degrees. Twenty three procedures were performed.

**Results**: The tests made it possible to bring the prototype forward on a variety of aspects, including reliability, steadiness, ergonomics and dimension. The ease of installation, which takes only five minutes and the easy handling of the robot made it possible for 21 out of 23 procedures to be performed without the need of an assistant. Voice recognition was chosen as the control system.

**Conclusion**: The LER robot is a robotised camera holder guided by the surgeon's voice that can eliminate the need for an assistant to hold the camera during laparoscopic surgery. The ease of installation and manufacture should make it an effective and inexpensive system in the lateral and dorsal supine positions. Ongoing randomized, prospective clinical trials will soon validate this robot prior to marketing.

**Introduction**

In laparoscopy, static camera-holding arms do exist, but do not allow the necessary changes in field of vision, particularly on zoom movements [1]. Light endoscope-holder robots were introduced to do away with the need for an assistant. AESOP®, EndoAssist ® and LapMan® are three of the systems currently on the market. They have been proven effective in ensuring a stable image and reducing operation times [2, 3] but have a limited distribution range, due to their large dimension and high price [4, 5]. The aim of TIMC-GMCAO Laboratory was to create a miniaturised light endoscope-holder robot, offered at a lower cost. We are presenting the results of the trials carried out with the LER prior to clinical assessment.

**Materials and Methods**

The LER (light endoscope robot) developed by TIMC-GMCAO Laboratory consists of a compact camera-holder robot (figures 1 and 2) resting directly on the patient's abdomen, and an electronic box containing the electricity supply and robot controllers [6-8]. The robot moves an optical trocar (tested diameter : 10 mm) making it possible guide and move the endoscope, the control of the zoom being included. Three miniaturised back-driveable motors are integrated. One motor is used to control the endoscope's insertion depth (motor 1) (figure 3), the second enables the core needle to rotate on an axis (motor 2) (figure 4), and the third enables endoscope pan-tilt (motor 3) (figure 5). It is possible to move the robot manually when the motors are off.

Gears and spoked wheels are used to control the rotations and a cable wound around a pulley device and held by a spring, makes it possible to control the endoscope's insertion depth (figure 2). The LER has three ranges of movement: 360° rotation around the vertical axis, an 80° tilt from the vertical position; and a horizontal incline of the endoscope 20 cm along its axis. A patent was filed in May 2002. A pedal, a microphone-equipped portable PC and vocal recognition software (Microsoft SpeechStudio®) are connected to the control box. Vocal recognition can be replaced by remote control. The robot is connected to the control box by two cables.

The compact architecture was designed so as to not hamper the surgeon's movement and enable rapid set-up, while keeping costs low[9]. A fully sterilisable robot place directly on the patient's abdomen was ultimately chosen to save time in putting together the robot.

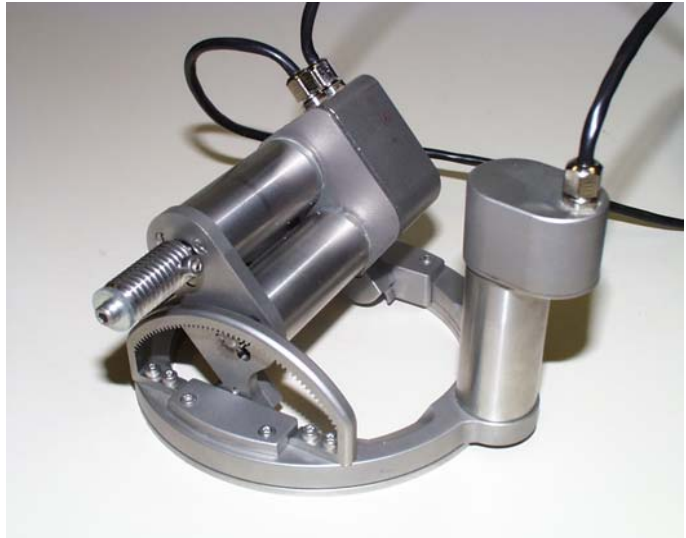

Figure 1: LER, or light endoscope robot

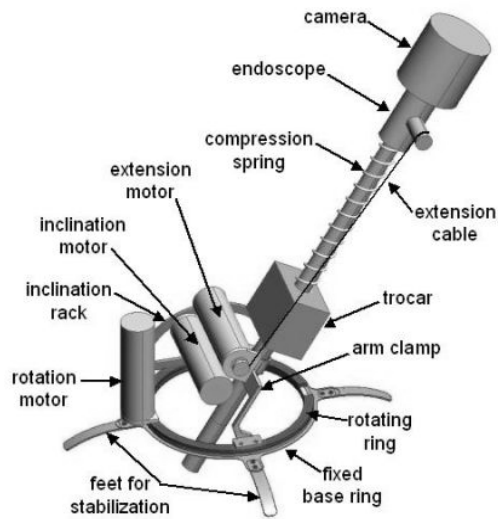

Figure 2: LER robot scheme

The robot is approximately 110 mm in diameter and 75 mm high. It weighs 625 g. Three prototypes have been produced in stainless steel. The architecture, based on a rotating circle, was the foundation for the development.

From September 2003 to January 2005, the LER's feasibility of use was studied on cadavers and laboratory pigs.

A number of developments on the LER were tested by 8 surgeons (6 urologists and 2 digestive surgeons.

The procedures performed on the cadavers included: 4 pelvic lymph node dissections, 2 right nephrectomies, 2 left nephrectomies, 1 right surrenalectomy, 1 left surrenalectomy, 2 prostatectomies, 2 cholecystectomies, 1 small bowel resection-anastomosis, and 3 appendicectomies. The procedures performed on the pigs included: 1 cholecystectomy, 1 splenectomy, 1 right nephrectomy, 1 left nephrectomy and 1 cystectomy. In total, 23 were performed on 10 cadavers and 2 laboratory pigs.

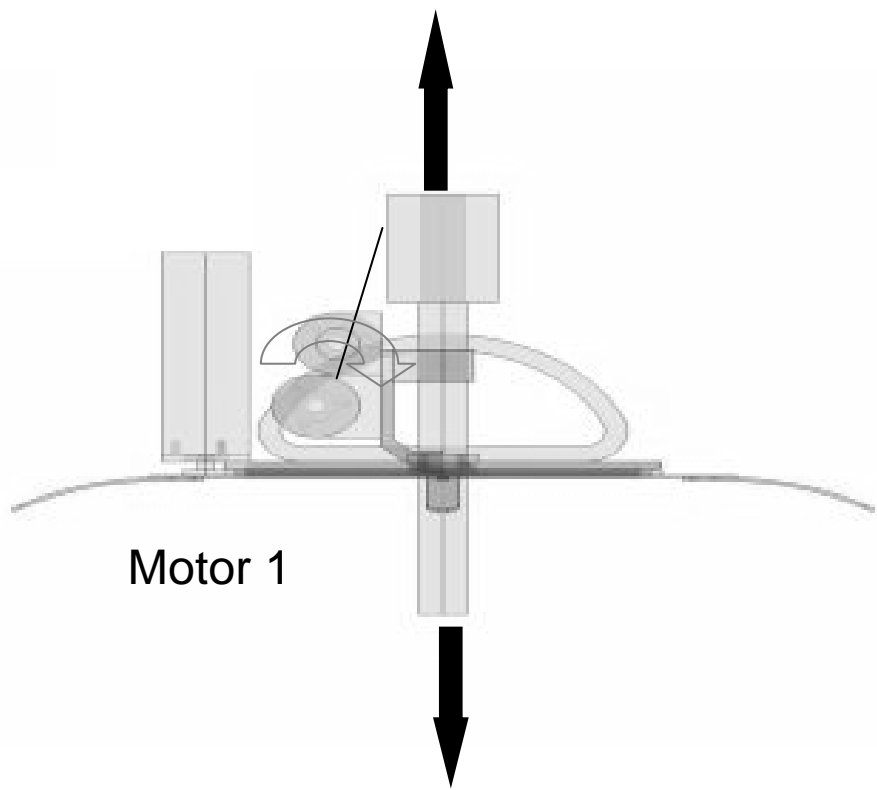

Figure 3: Action of Motor 1 enabling zoom movements as the cable winds around the pulley. The return to the highest position is enabled by a spring.

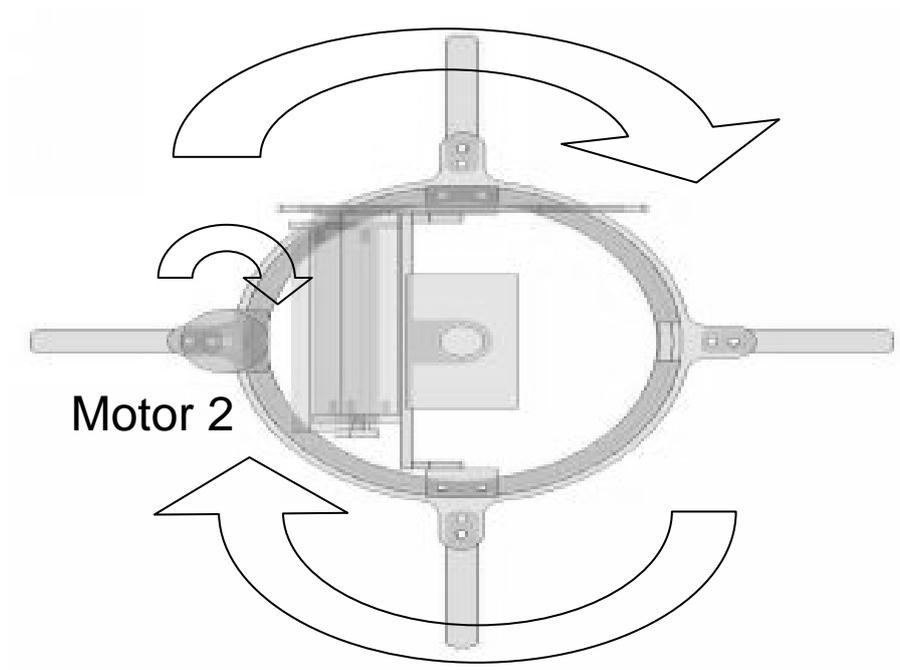

Figure 4: Encoscope rotation movements around trocar axis (motor 2)

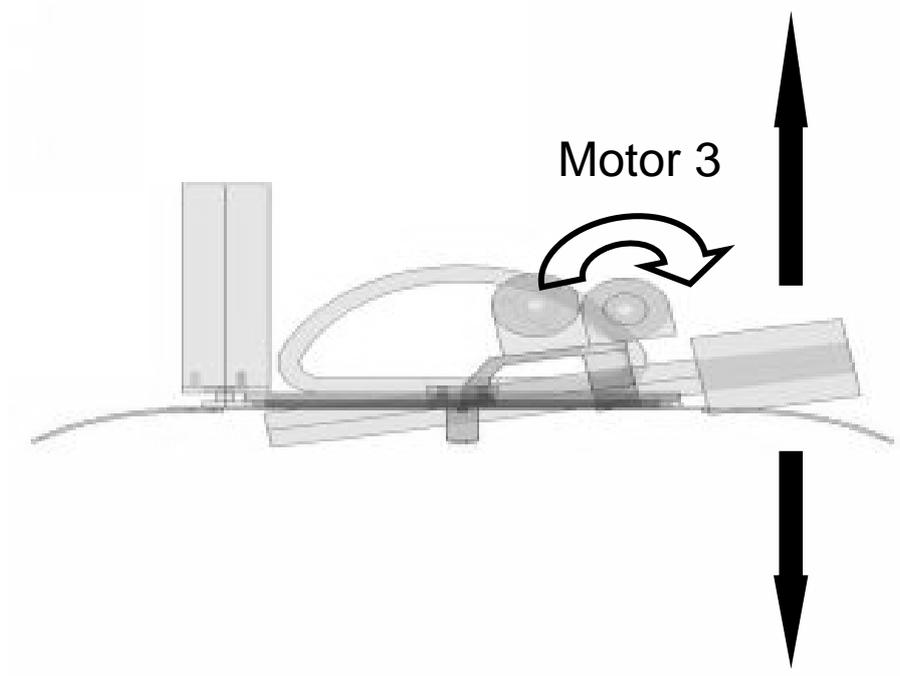

Figure 5: Up-down movements using Motor 3

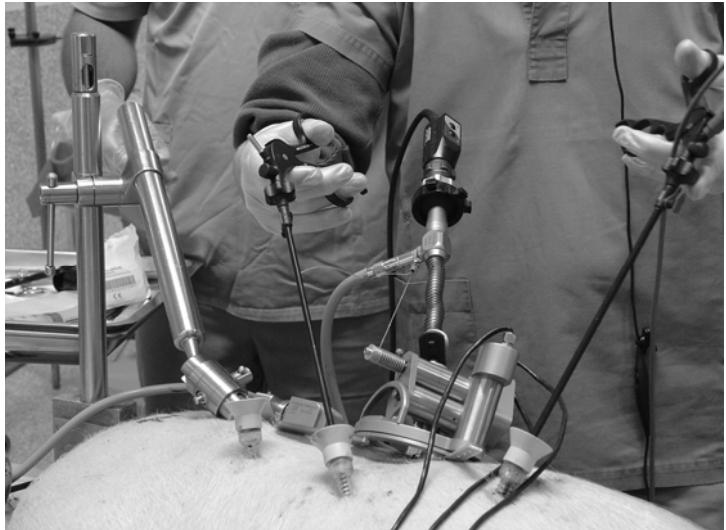
Figure 6: Robot firmly set

## Results

In two procedures, it was necessary to stop the robot and resume using camera control because of two robot breakdowns: the base ring came out of position; and once one motor overheated. The accidents were corrected on the following version of the LER.

- *Robot steadiness:*

The robot had to be fixed to the operating table using an articulated arm (figure 6). The tests on the pelvi-trainer had given hope that it would be possible to lay the robot on the patient, but after several hours of surgery, the device left marks on the animals' skin. Clamping the robot made it possible to correct that problem, while also improving the stability of the image. The articulated arm does present a drawback in that it can extend over or near the operating area. As a result, it is necessary to reflect on how to position the LER before making an incision, so that the set-up does not hinder the movement of the instruments.

- *Command interface:*

The interface customarily used was the voice recognition system, as it proved reliable in recognising orders, even with surgeons who had no experience using the robot. The sound environment, which was particularly noisy at times, did not create any problems, as described for other systems [5]. Training was preferable beforehand so as to ensure that the software would recognise a specific voice.

Manual commands using a mini-keypad clipped onto an instrument can also be used with high precision and good ergonomics. However, the need to move the keypad onto another instrument when changing tools proved a hindrance.

- *Set-up and removal time:*

The quick and easy set-up for the robot varied from 5 to 40 minutes, as different systems were tested, and depending on experience. During the last procedures, the set-up times were the shortest. The positioning of the articulated arm used to clam the robot had to be given consideration before the procedure, so that it would not affect the positioning of the trocars. Removal was very quick and a test conversion to laparotomy on a laboratory pig, following healing of the splenic pedicule during a splenectomy, showed that the system can be easily removed in 30 seconds.

- *Magnitude of intra-abdominal field of vision:*

The 360° movement made it possible to explore the entirety of the abdominal cavity. The robot's speed of movement was satisfactory (75° per second, 80 mm/s), making it possible to focus on a point of interest quickly in continuous mode. The discontinuous movement function made it possible to change the angle of vision by just a few degrees, thereby enabling small, finer readjustments during already-meticulous movements. We were hampered by the inadequate insertion depth of the endoscope during a uretro-vesical anastomosis, performed as part of radical prostatectomy. The lack of zoom to the very end of the pelvis was due to the length of the compact spring, which reduced the opportunity to move the endoscope in depth (figure 7). This problem has been resolved.

- *Comfort of use*

The surgeons using the robot were satisfied with the ergonomics and comfort of the LER.

During the experiments, the standard headset was replaced by a bluetooth headset to reduce the amount of cables.

- *Robot dimensions and trocar positioning*

The advantage with the LER lies in its compactness. The assistants were not hampered by the robot when setting up or moving around during the procedure. The clamp did not hinder the surgeon's movements either. As regards the position of the trocars, the 11-cm-diameter base did not require any change in trocar position during the procedures that we performed using laparoscopy. In contrast, in lateral supine position, during the first procedure, we were hampered by the clamp arm, which had been placed on a trocar insertion point.

- *Position of patient.*

The initial robot had been created for use in dorsal supine position (figure 8). The need to clamp the robot extended the potential patient positions to include the lateral supine position (figure 9). The stability of the whole system was satisfying. The diameter of the robot's base made it uneasy to introduce the trocars during a lomboscopy, which does not appear to be an indication. With the transperitoneal approach, we did not encounter any problems. The trials on animals and cadavers were compelling in this position.

The set up in dorsal supine position did not raise any particular problems.

- *System Reliability*

Reliability was the main problem during the trials, but has since been resolved. 2 motors had to be changed as they had deteriorated following an electrical trial.

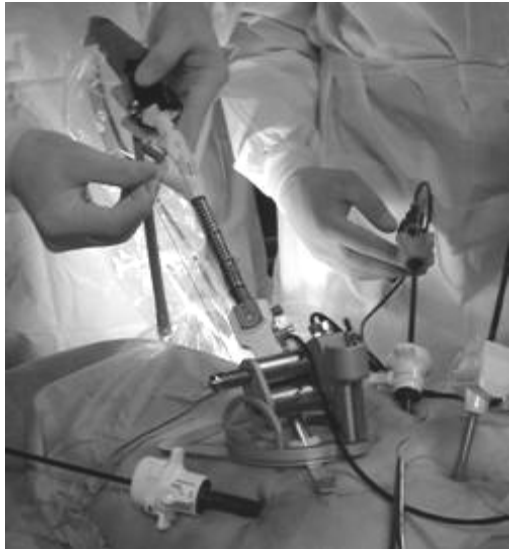
Figure 7: Zoom (spring and cable)

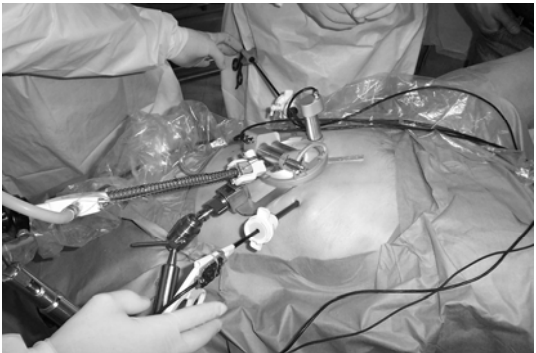
Figure 8: Dorsal supine position

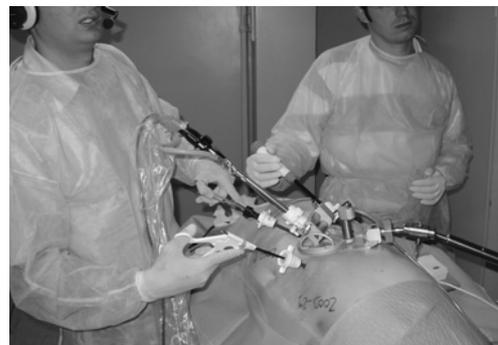
Figure 9: left lateral supine position

The first robot sterilisation tests showed fragility in the cable sheaths, which had to be replaced by silicone sheaths. Sterilisation was performed by soaking the cables in an aldehyde-free detergent bath for 15 minutes, then sterilising them in an autolave at 134° for 18 minutes.

**Discussion**

Through use of the LER, the robot proved reliable on cadavers and laboratory pigs. The reliability problems that appeared were solved during the final tests. The patient security issues appear to have been overcome and it appears possible to set the robot directly on the patient. The surgical procedures were never forced to deviate from conventional laparoscopy. The most noteworthy benefits were the stability of the useful image during anastomosis and the possibility of operating alone. However, this required robot set-up times ranging from 5 to 40 minutes, depending on the team's level of experience.

The recognised benefits of endoscope-holder robots are the ability to do without an aid, the image stability, the lesser degree of fatigue for the surgeon and the lower frequency of camera soiling[10]. However, their large dimensions and, above all, their price, have slowed down their spread. The aim is to produce an affordable system that can be cost-effective, as it does away with the need for an operating assistant and enables quick movements. For this reason, it is necessary that the system be set up quickly, as it appears difficult that the robot should be able to provide any improvement in procedure time compared to an assistant-facilitated procedure, even if the literature does report lower times with EndoAssist [4] and AESOP[10].

It is difficult to estimate in advance the price of the LER, as the price of a prototype does not include the development and marketing costs.

The contribution of the LER as compared to existing systems lies above all in its compact nature, which makes it possible not to clutter the operating table and room area, while not hampering the assistant's movements. Trocar position was not hampered by the robot's base diameter during laparoscopy.

Feasibility has been affirmed by the pre-clinical tests and many changes have taken place. Its ability to withstand time and repeated sterilisation procedures is yet unknown.

## Conclusion

The LER is a light endoscope-holder robot with miniature design, set on the patient's abdomen, in order to make the system less cumbersome and costly than the robots currently on the market.

Pre-clinical trials have shown that LER use is feasible in laparoscopic digestive and urological surgery on cadavers and animals.

Clinical trials will tell whether the system is suited to man.